\documentclass[conference]{IEEEtran}
\usepackage{alltt                                    
          , multirow
          , booktabs
          , listings
          , graphicx
          ,float
	,cite
          ,verbatim
         ,mathtools
	,url
	,amsmath
}
\usepackage[table]{xcolor}
\usepackage[numbers]{natbib}     
\usepackage{syntax}
\usepackage{algorithm}
\usepackage{enumitem}
\usepackage{threeparttable}
\usepackage{algpseudocode}
\usepackage{pbox}
\usepackage{balance}
\usepackage{hhline}

\usepackage{expl3}
\ExplSyntaxOn
\newcommand\latinabbrev[1]{
  \peek_meaning:NTF . {
    #1\@}%
  { \peek_catcode:NTF a {
      #1., \@ }%
    {#1., \@}}}
\ExplSyntaxOff

\newcommand{\CASE}[1]{\STATE \textbf{case} #1\textbf{:} \begin{ALC@g}}
\newcommand{\ENDCASE}{\end{ALC@g}}

\newcommand{\DEFAULT}{\STATE \textbf{default:} \begin{ALC@g}}
\newcommand{\ENDDEFAULT}{\end{ALC@g}}
\newcommand{\DEFAULTLINE}[1]{\STATE \textbf{default:} }
\let\footnotesize\scriptsize

\colorlet{shadecolor}{gray!40}

\newsavebox{\supbox}
\newcommand{\bsup}{\begin{lrbox}{\supbox}$\tt\scriptstyle}
\newcommand{\esup}{$\end{lrbox}{}^{\usebox{\supbox}}}
\def\eg{\latinabbrev{e.g}}
\def\ie{\latinabbrev{i.e}}

\algnewcommand{\LineComment}[1]{\State \(\triangleright\) #1}

\definecolor{lightpurple}{rgb}{0.8,0.8,1}
\definecolor{codebg}{RGB}{255,255,255}
\definecolor{commentcolor}{RGB}{11,140,11}
\begin{document}
%

\title{Predicting Usefulness of Code Review Comments using Textual Features and Developer Experience\vspace{-0.3cm}}

%
%
%
%


\author{\IEEEauthorblockN{Mohammad Masudur Rahman  ~~~ Chanchal K. Roy~~~$^\S$Raula G. Kula}
\IEEEauthorblockA{University of Saskatchewan, Canada, $^\S$Osaka University, Japan\\
\{masud.rahman, chanchal.roy\}@usask.ca, $^\S$raula-k@ist.osaka-u.ac.jp}
}

\maketitle
\begin{abstract}
Although peer code review is widely adopted in both commercial and open source development, existing studies suggest that such code reviews often contain a significant amount of non-useful review comments.
Unfortunately, to date, no tools or techniques exist that can provide automatic support in improving those non-useful comments.
In this paper, we first report a comparative study between useful and non-useful review comments where we contrast between them using their textual characteristics, and reviewers' experience.
Then, based on the findings from the study, we develop RevHelper, a prediction model that can help the developers improve their code review comments through automatic prediction of their usefulness
during review submission.
Comparative study using 1,116 review comments suggested that useful comments share more vocabulary with the changed code, contain salient items like relevant code elements, and their reviewers are generally more experienced. 
Experiments using 1,482 review comments report that our model can predict comment usefulness with 66\% prediction accuracy which is promising.

Comparison with three variants of a baseline model using a case study validates our empirical findings and demonstrates the potential of our model.
\end{abstract}



\IEEEpeerreviewmaketitle

\section{Introduction}\label{sec:introduction}
Peer code review has been considered as one of the best engineering practices in software development for over last 35 years \cite{expect,convergent}. 
It works as a quality safeguard for the patches (\ie\ software code changes) that are to be integrated into the master codebase of a software system.
It detects simple source code defects (\eg\ logical errors) and coding standard violations (\eg\ lack of descriptive identifier names) in the early phases of the development, which reduces the overall cost \cite{shane-quality}.
Despite the proven benefits, the formal code inspection (also called Fagan's inspection) was never widely adopted due to its high costs and inapplicability to distributed software development (\ie\ demands synchronous meeting among developers) \cite{expect}.
Interestingly, \emph{modern code review}, a lightweight version of the peer code review that is assisted with various tools, has gained remarkable popularity in recent years both in the industry (\eg\ Microsoft, Google) and in the open source development (\eg\ Android, LibreOffice) \cite{expect}.
The main building blocks of modern code review are 
the associated review comments provided by the reviewers during code review \cite{useful}.
Such comments are generally consulted by the patch submitters before correcting their code changes and then submitting the next patch. 
Hence, the quality of the review comments is crucial, and the comments are expected to be accurate, thorough \cite{devsee} and useful \cite{useful}.
Inaccurate, ambiguous or low quality review comments could mislead the submitters, kill valuable development time unnecessarily, and thus, could simply defeat the purpose of the code review. 
A recent study \cite{useful} reported that 34.50\% of the code review comments from five major projects of Microsoft are \emph{non-useful}.  
Automatic filtration of those review comments could have saved the development time of the patch submitters \cite{useful}. 
Alternatively, meaningful insights on the quality (\eg\ prediction of usefulness) of those comments accompanied by helpful suggestions could have assisted the reviewers in improving their comments in the first place. 
Unfortunately, to date, no tools or techniques exist that predict usefulness of the review comments during the submission of a code review. 

In fact, \citet{useful} first explore professional developers'  perceptions on the usefulness of code review comments using a qualitative study, and identify several characteristics of the useful and non-useful review comments at Microsoft.   
According to them, developers find those review comments \emph{useful} that trigger code changes within the vicinity (\ie\ 1--10 lines) of the comment locations. 
\citet{devsee} conduct a similar study involving professional developers from open source domain--\emph{Mozilla}, and analyze the perceptions of the developers on the quality of code reviews.  
They suggest that the quality of code reviews is primarily associated with the thoroughness (\ie\ completeness) of review feedback, \ie\ comments.
While both studies offer insights on the usefulness of review comments by employing qualitative methods, 
there has been a marked lack in the application of such insights in practical settings using automated approaches.
Although \citeauthor{useful} develop a classifier model for review comments, their feature set is constrained, and many of their features are not  available during review comment submission. Thus, the model is not capable of predicting usefulness of the new review comments to be submitted.
On the other hand, we adopt their definition of \emph{usefulness of comments} \cite{useful} and provide automated supports for identifying and thus improving 
the non-useful review comments during review submission.
Besides, given the  earlier findings \cite{useful,devsee}, textual features of the review comments are not properly studied yet which might be linked to their usefulness. 
\begin{figure*}[!t]
\centering
\includegraphics[width=7.1in ]{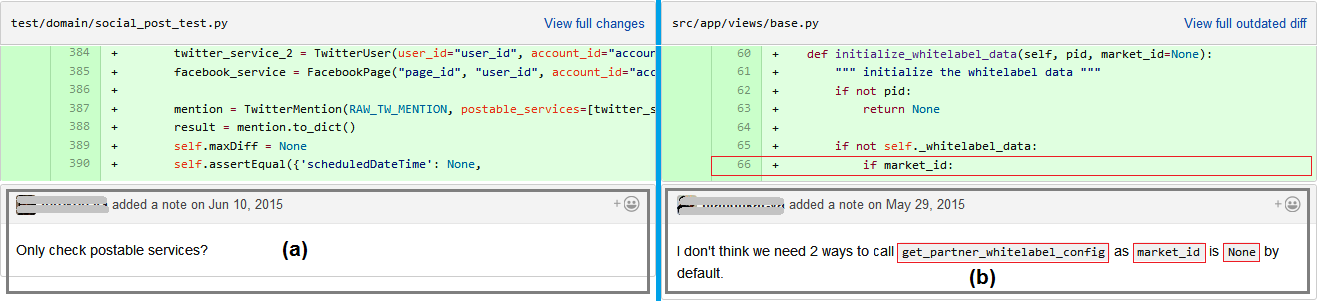}
\vspace{-.7cm}
\caption{Review comments from \emph{ABC} code reviews--(a) Non-useful review comment, and (b) Useful review comment}
\label{fig:motiv}
\vspace{-.5cm}
\end{figure*}

In this paper, we (1) report a comparative study on the usefulness of code review comments using their textual features and reviewers' expertise, and then (2) propose a prediction model --\emph{RevHelper}-- based on the findings from our study. 
We first motivate our idea using a comparative study between useful and non-useful review comments, and answer two research questions. 
Review comments are marked as \emph{useful} or \emph{non-useful} based on the suggested heuristics of \citeauthor{useful} (\eg\ change-triggering capability) that are derived from an interview with professional developers from Microsoft. 
Then, we develop a machine learning model for  predicting usefulness of a review comment based on the perceived signals from the comparative study. 
Such a model not only provides an approximate view on the quality of review comments, but also can assist the code reviewers in improving their review comments, especially the non-useful ones, through meaningful insights and complementary information. 

\begin{figure}[!t]
\centering
\includegraphics[width=3.45in ]{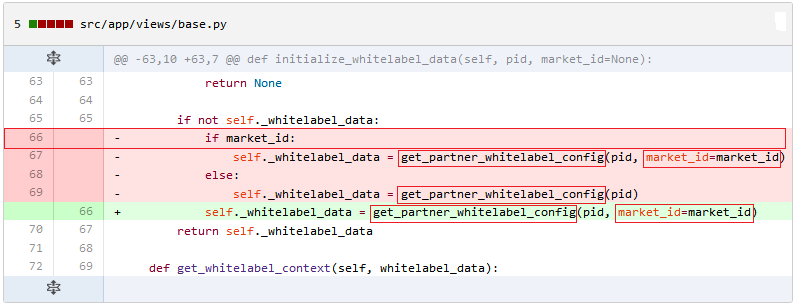}
\vspace{-.6cm}
\caption{Code change triggered by the useful review comment from Fig. \ref{fig:motiv}-(b)}
\label{fig:cchange}
\vspace{-.6cm}
\end{figure}

For example, the review comment in Fig. \ref{fig:motiv}-(a)(bottom row)--\emph{``Only check postable services?"}--did not trigger any code change, and the patch submitter also found the comment non-useful \cite{useful}. We note that the comment  
does not warrant any clear action (\ie\ change of code) rather asks for a clarification.  
On the other hand, the review comment in Fig. \ref{fig:motiv}-(b)(bottom row)--\emph{``I don't think we need 2 ways to call} \texttt{get_partner_whitelabel_config} \emph{as} \texttt{market_id} \emph{is} \texttt{None} \emph{by default"}-- triggered a code change within its vicinity (\ie\ next line, Fig. \ref{fig:cchange}), and the patch submitter also found the comment useful.
Please note that the comment contains relevant code artifacts (\eg\ \texttt{market_id}), and also suggests a code change on invoking the method-- \texttt{get_partner_whitelabel_config()}.  
Such a warranted action along with references to the artifacts or concepts of interest from code
could have led the review comment to be useful. In our research, we study several of such textual features of the review comments along with the reviewers' expertise, contrast between useful and non-useful comments, and then capture their essence into a usefulness prediction model--RevHelper.
To the best of our knowledge, ours is the first study that not only contrasts between useful and non-useful comments extensively (\ie\ using two dimensions and eight independent features) but also automatically predicts usefulness of the review comments during code review submission. 
The major focus of the earlier studies \cite{useful,devsee} was to analyze the review comments qualitatively by involving professional developers. 
The datasets used for our comparative study and the experiments are hosted online \cite{revhelper} for replication purposes.

Our investigation using 1,116 code review comments from four commercial subject systems of \emph{ABC Company} (blinded for double-blind reviewing) reported that (1) 44.47\% of those comments were found non-useful (\ie\ did not trigger any code changes \cite{useful}), 
(2) useful review comments share more vocabulary with the changed code, and contain more relevant code elements (\eg\ program entity names) and relatively less stop words, and (3) reviewers' past experience positively influence the usefulness of their comments.
Experiments using 1,116 code review comments from \emph{ABC Company} (\ie\ ground truth is based on change-triggering heuristic of \citeauthor{useful}) suggest that our model can predict comment usefulness with 66\% overall accuracy, 66\% precision and 66\% recall which are promising given that ours is the first complete model of its kind. 
Finally, a case study using 366 new review comments and comparison with three variants of a baseline model \cite{useful} not only validates our empirical findings
but also demonstrates the potential of our prediction model.
Thus, we make the following contributions in this paper.
\begin{itemize}[noitemsep,topsep=0pt]
\item  A comparative study that contrasts between useful and non-useful review comments using five textual characteristics and three reviewers' experience features.
\item  A novel and promising model for predicting the usefulness of a new review comment during review submission. 
\item  Comprehensive evaluation of the proposed model, and validation using a separate case study and three variants of a baseline model from the literature.
\end{itemize}

\begin{table*}[!t]
\centering
\caption{Overview of the Variables and Dimensions Used}\label{table:independent}
\vspace{-.2cm}
\resizebox{5.1in}{!}{%
\begin{threeparttable}
\begin{tabular}{l|l|l}
\hline
\textbf{Independent Variable} & \textbf{Dimension} &  \textbf{Description}\\
\hline
\hline
Reading Ease (RE) \cite{low,qualityquestion} & textual & numerical estimation of ease with which a comment can be read \\
\hline
Stop Word Ratio (SWR) \cite{wordsim} & textual & frequently occurring words in natural language texts with little or no semantics \\
\hline
Question Ratio (QR) \cite{useful} & textual & percentage of interrogative sentences in each review comment\\  
\hline
Code Element Ratio (CER) \cite{explanation} & textual & percentage of source code related tokens in each review comment \\
\hline
Conceptual Similarity (CS) \cite{antoniol} & textual & lexical similarity between changed source lines and each review comment \\
\hline
Code Authorship (CA) \cite{useful,ownership} & experience & number of commits authored on a file by a developer \\ 
\hline
Code Reviewership (CR) \cite{useful,ownership} & experience & number of commits on a file reviewed by a developer \\
\hline
External Library Experience (ELE) \cite{correct} & experience & percentage of external libraries a developer familiar with from a reviewed file \\

\hline
\end{tabular}
\centering
\end{threeparttable}
}
\vspace{-.5cm}
\end{table*}

\section{Comparative Study between Useful and Non-useful Review Comments}
Given that existing findings are mostly derived from qualitative studies involving software developers \cite{useful,devsee}, we are interested to further understand the topic of usefulness of review comments from a comparative perspective. In particular, we want to find out what items are missing from the non-useful comments or what distinguishing features of the useful comments are leading them to be useful.
Hence, we conduct a comparative study using 1,116 code review comments taken from four commercial subject systems of \emph{ABC Company}. 
In our study, we answer the two research questions as follows.
\begin{itemize}[noitemsep,topsep=0pt]
\item  \textbf{RQ$\mathbf{_1}$}: Are useful review comments different from non-useful comments in terms of their textual properties? 
\item  \textbf{RQ$\mathbf{_2}$}: How does experience of the developers help them write useful review comments during a code review? 
\end{itemize}

\subsection{Dataset for Study}
\label{sec:dataset}
\textbf{Dataset Collection:} We collect 1,200 recent code review comments for our study from four commercial subject systems of \emph{ABC Company}.
In GitHub code reviews, reviewers generally submit two types of review comments--\emph{in-line comments} and \emph{pull request comments}. While an in-line comment refers to a particular code change directly, the pull request comment targets the whole collection of patches submitted in the request. 
In this research, we focus on \emph{in-line} comments of a code review as was also done by \citet{useful}.
We collect 300 most recently posted \emph{in-line} review comments from each of the four subject systems by accessing GitHub API using a client library--\emph{github-api} \cite{githubapi}. Once collected, we request the company developers for initial feedback on their comments. The company reported that few of the pull requests were not originally intended for merging commits rather than for scaffolding purposes. We thus discarded such pull requests along with their review comments from our dataset, and finally ended up with a total of 1,116 review comments.    

\textbf{Annotation of Review Comments:} We manually annotate each of the collected comments as either \emph{useful} or \emph{non-useful} by
applying \emph{change triggering} heuristic of \citet{useful}.
That is, if any comment triggers one or more code changes within its vicinity (\ie\ 1--10 lines) in the subsequent commits (\ie\ patches) of a pull request, the comment is marked as \emph{useful}. On the other hand, if the comment either triggers code changes in unrelated places or does not trigger any change at all, the comment is marked as \emph{non-useful}. 
Not only the heuristic was recommended by the professional developers from \emph{Microsoft} but also it was found highly effective for comment classification \cite{useful}, which justify our choice of the heuristic for the manual annotation.
The whole annotation task took about 20 man-hours.
We found 55.53\% of our collected comments  as \emph{useful} and 44.47\% of them as \emph{non-useful}.
Table \ref{table:dataset} shows the details of our dataset used for the study.


\subsection{Determining Independent Variables and Dimensions}\label{sec:independent}
According to \citet{devsee}, a significant fraction (\ie\ 38\%) of the developers consider that \emph{clear} and \emph{thorough feedback} is the key attribute of a well-done code review.
The clarity of any natural language artifact can be determined based on a popular metric called \emph{readability} as was also applied by several existing studies \cite{low,msrch2015masud,qualityquestion}.
The thoroughness of a comment might be linked to the conciseness and salience of the comment texts \cite{autocomment}.
\citet{useful} report positive impact of developers' prior experience upon their code reviews.
\citeauthor{devsee} also suggest that having enough domain knowledge by the reviewers is a prerequisite for effective code reviews. 
Such experience can be estimated by using their code authorship, review activities \cite{useful} or even familiarity with the external libraries used in the changed source files under code review \cite{correct}. 
Given that many of these findings on code reviews are based on qualitative studies, we want to revisit them using a comparative study between useful and non-useful comments.
We thus consider a total of eight independent variables  from two dimensions and one response variable--\emph{usefulness of comment} for our study.
Table \ref{table:independent} provides an overview of our selected independent variables where first five are related to  \emph{textual properties} of the review comment and the remaining three are proxies for \emph{developer's experience}.
For each variable, we present our method for comparison between useful and non useful comments. We apply several statistical tests to determine the differences between the both groups.


\begin{table}[!t]
\centering
\caption{Dataset for Comparative Study}\label{table:dataset}
\vspace{-.2cm}
\resizebox{3.2in}{!}{%
\begin{threeparttable}
\begin{tabular}{l|c|c|c|c}
\hline
\textbf{System} & \textbf{Pull Requests} &  \textbf{Useful Comments}  & \textbf{Non-useful Comments} & \textbf{Total}\\
\hline
\hline
CS & 80 &  153 (59.77\%) & 103 (40.23\%) & 256 \\
\hline
SM & 97 &  164 (58.36\%)  & 117 (41.64\%)    & 281 \\
\hline
MS & 88 &  155 (53.82\%) &133 (46.18\%)  & 288 \\
\hline
SR & 101 & 146 (50.17\%) & 145 (49.83\%)  & 291 \\
\hline 
\hline
\textbf{Total} & 366 &  618 (55.53\%) & 498 (44.47\%) & 1,116\\
\hline
\end{tabular}
\centering
\end{threeparttable}
}
\vspace{-.3cm}
\end{table}

\begin{figure}[!t]
\centering
\includegraphics[width=2.6in ]{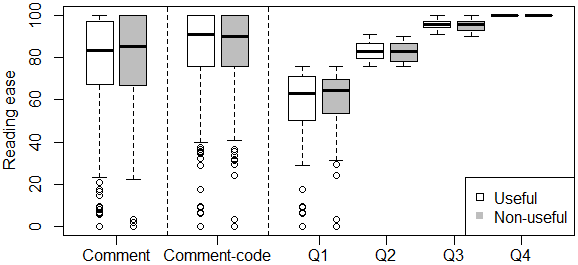}
\vspace{-.3cm}
\caption{Reading ease of code review comments (\textbf{Q1}=0.25 quantile, \textbf{Q2}=0.50 quantile, \textbf{Q3}=0.75 quantile, and \textbf{Q4}=1.00 quantile)}
\label{fig:rease}
\vspace{-.6cm}
\end{figure}



\subsection{Textual Features}\label{label:textual}
\textbf{Reading Ease:}\label{sec:rease}
Flesch-Kincaid reading ease is one of the most popular metrics for determining readability of any regular texts \cite{low,qualityquestion}. It is based on a ranking scale from 0 to 100 where the higher value represents lower complexity of the texts and vice versa.   
Fig. \ref{fig:rease} summarizes our findings on reading ease.
We see that the median ease is between 80 and 90 which suggests that both comment types are easy to read.
Unfortunately, unlike the qualitative findings suggest \cite{devsee}, we did not find any significant difference in the reading ease (\ie\ \emph{Mann-Whitney Wilcoxon test, p-value=0.32$>$0.05, Cohen's D=0.04$<$0.20}) between useful and non-useful comments.
We then manually analyze the comments, and found that comment texts contain both regular texts and structured code elements (\eg\ identifier names, code fragments).
We discard the code elements from the comments using appropriate regular expressions, and calculate the reading ease again.
As Fig. \ref{fig:rease} shows, we found that overall reading ease significantly increased for both useful (\ie\ \emph{p-value=0.00, Cohen's D=0.28}) and non-useful comments (\ie\ \emph{p-value=0.00, Cohen's D=0.25}) with medium effect sizes.
We note that useful comments are a bit easier to read than non-useful comments with only regular texts considered.
However, their reading ease difference is not statistically significant (\ie\ \emph{p-value=0.80$>$0.05, D=0.00}).  

We also divide the reading ease measures for each of the comment types into four quartiles-- Q1, Q2, Q3 and Q4, and then compare them using box plots, as done by relevant studies \cite{nontechnical}.
However, the readability difference between useful and non-useful comments was not found statistically significant (\ie\ \emph{p-values$>$0.05}) for any of the four quartiles.

\begin{figure}[!t]
\centering
\includegraphics[width=2.6in ]{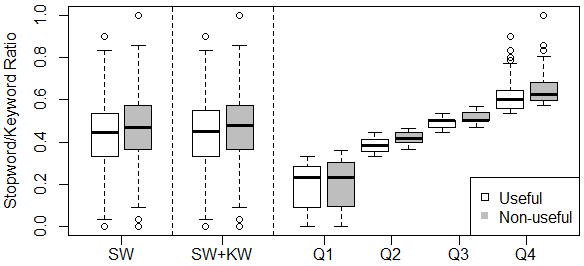}
\vspace{-.3cm}
\caption{Stop word and keyword ratio (\textbf{SW}=Stop words, and \textbf{KW}=Keywords, \textbf{Q1}=0.25 quantile, \textbf{Q2}=0.50 quantile, \textbf{Q3}=0.75 quantile, and \textbf{Q4}=1.00 quantile)}
\label{fig:sratio}
\vspace{-.6cm}
\end{figure}

\textbf{Stop Word Ratio:}
Stop words are frequently occurring words in the regular texts with very little or no semantics \cite{wordsim}. 
Given the limited size of a review comment, 
we are interested to find out if there exists any distributional difference of stop words between useful and non-useful comments.
We determine stop word ratio over all words from each comment, and compare the ratios from both comment types. 
We use a standard list \cite{stopword} of stop words  for our analysis.
Fig. \ref{fig:sratio} summarizes the comparative analysis between both types.
We see that the median stop word ratio is relatively higher for non-useful review comments than that of useful comments. 
In fact, the ratio difference is statistically significant (\ie\ \emph{p-value=0.01}$<$0.05) although the effect size is small (\emph{Cohen's D=0.14}).
This partially supports our conjecture that non-useful comments contain relatively less meaningful words which might have reduced their usefulness.  

Programming keywords can be considered as an equivalent to stop words in the source code \cite{refoqus}. 
Since review comments sometimes contain source code elements along with regular texts, 
we recompute the stop word and keyword ratios (\ie\ of Python language) and compare the ratios for both comment types.
As Fig. \ref{fig:sratio} shows, the ratio is still higher for non-useful comments than that of useful comments, and their difference is statistically significant (\emph{p-value=0.01}$<$0.05). 
However, the effect size of the difference is still small (\emph{Cohen's D=0.14}). 
We also divide the stop word ratios for each comment type into four quartiles, and compare between useful and non-useful comments.
While the difference for Q1 is not significant (\emph{p-value=0.55$>$0.05}), we found statistically significant difference for Q2 (\ie\ \emph{p-value=0.00, Cohen's D$>$1.00}), Q3 (\ie\ \emph{p-value=0.00, Cohen's D=0.89}) and Q4 (\ie\  \emph{p-value=0.00, Cohen's D =0.54}) with medium to large effect sizes, which are interesting, and support our conjecture.

\begin{figure}[!t]
\centering
\includegraphics[width=2.4in ]{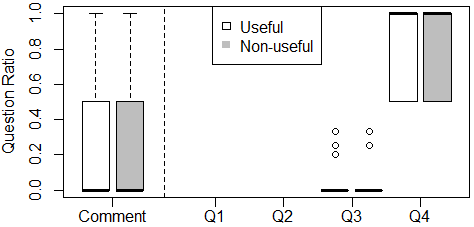}
\vspace{-.3cm}
\caption{Question ratio of review comments (\textbf{Q1}=0.25 quantile, \textbf{Q2}=0.50 quantile, \textbf{Q3}=0.75 quantile, and \textbf{Q4}=1.00 quantile)}
\label{fig:qratio}
\vspace{-.6cm}
\end{figure}

\textbf{Question Ratio:}
Reviewers often ask clarification questions during code review. \citeauthor{useful} suggest that such questions might be helpful for knowledge dissemination among the developers but do not improve the quality of a submitted patch.
Thus, review comments as a form of clarification questions are not considered useful by the professional developers \cite{useful}.
We revisit that proposition in our study, and investigate if that is supported by our collected data or not.
In particular, we want to see if there exists a difference in the ratio of interrogative sentences over all sentences from a review comment between useful and non-useful comments.  
We thus determine the number of questions from each comment using appropriate regular expression, and compare the question ratio between both comment types. 
Fig. \ref{fig:qratio} summarizes our comparative findings. We see that about 50\% of the review comments from both categories do not contain any questions (\ie\ median question ratio = 0). More importantly, the distributions of question ratio of both comment types look similar, and we did not find any statistically significant difference (\ie\ \emph{p-value=0.08$>$0.05, Cohen's D=0.12}) between them.
An investigation using quartile analysis also revealed the same finding.  
Finally, a Kruskal-Wallis test confirmed that the distribution of question ratio for both comment types are identical (\ie\ \emph{p-value=0.08$>$0.05}), and the usefulness of a review comment is not influenced by such distribution. 

\textbf{Code Element Ratio:}\label{sec:ceratio}
Since our investigation suggests that code segments or code-like elements in the review comments could affect their readability (\ie\ reading ease) significantly, we are interested to find out whether they also affect the usefulness of a comment or not.
Our interest also stems from the fact that the usefulness of review comments is determined based on their code change triggering capabilities in practice \cite{useful}. 
\citet{explanation} also report that a good code explanation in Stack Overflow posts generally overlaps (\ie\ shares artifacts) with the code snippet.
We thus believe that presence of relevant code elements in the review comment might improve its chance of being useful to the patch submitters during change implementation.
A comparative analysis between useful and non-useful comments on the code elements can offer us further meaningful insights. 
We thus collect all code segments and code-like elements from each of the review comments using a set of custom regular expressions, and discard false positives through a careful manual analysis. 
We then perform the comparative analysis where we investigate how frequently and how extensively the code elements are used in both types of review comments. 

\begin{table}[!t]
\centering
\caption{Source Code Elements in Review Comments}\label{table:ce}
\vspace{-.2cm}
\resizebox{3.5in}{!}{%
\begin{threeparttable}
\begin{tabular}{l|c|c|c|c|c|c|c}
\hline
\textbf{System} & \textbf{All} & \multicolumn{2}{c|}{\textbf{Useful Comments}} & \multicolumn{2}{c|}{\textbf{Non-useful Comments}} & \multicolumn{2}{c}{\textbf{Both}}  \\
\hline
& &  \textbf{CC} & \textbf{CWC}& \textbf{CC} & \textbf{CWC} & \textbf{CC} & \textbf{CWC}  \\
\hline
\hline
CS & 256 & 94 (61.44\%)  & 59 & 46 (44.66\%)  & 57 & 140 (54.69\%)  & 116 \\
\hline
SM & 281 & 87 (53.05\%) & 77 &  42 (35.90\%) & 75  & 129 (45.91\%)  & 152 \\
\hline
MS& 288 &  91 (58.71\%) & 64 & 70 (52.63\%)  & 63 & 161 (55.90\%)  & 127 \\
\hline
SR & 291 & 66 (45.21\%) &  80 & 77 (53.10\%) & 68 & 143 (49.14\%) & 148 \\
\hline
\textbf{Total}& \textbf{1,116} &\textbf{338 (54.69\%)} &  & \textbf{235 (47.19\%)} & & \textbf{Avg=51.34}\%& \\
\hline
\end{tabular}
\centering
\textbf{CC}=Review comments with code elements, \textbf{CWC}=Review comments without code elements 
\end{threeparttable}
}
\vspace{-.3cm}
\end{table}

From Table \ref{table:ce}, we see that about 51\% of all review comments contain one or more source code elements which is a significant amount.
We also note that code elements are more frequently used in the useful review comments than in the non-useful comments for each of the four systems except one (\ie\ SR).   
For example, about 45\%--61\% of the useful comments from our selected subject systems refer to at least one code element whereas such ratios reach up to 36\%--53\% for the non-useful comments. However, our statistical analysis suggested that these ratios are not significantly different (\ie\ \emph{Paired t-test, p-value=0.27$>$0.05}) between both comment types.


\begin{figure}[!t]
\centering
\includegraphics[width=2.3in ]{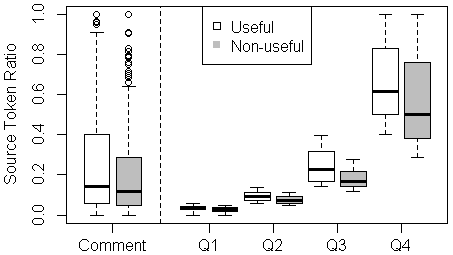}
\centering
\vspace{-.3cm}
\caption{Source token ratio of review comments (\textbf{Q1}=0.25 quantile, \textbf{Q2}=0.50 quantile, \textbf{Q3}=0.75 quantile, and \textbf{Q4}=1.00 quantile)}
\label{fig:ce}
\vspace{-.6cm}
\end{figure}

We also consider the extent to which code elements are used within each review comment, and investigate if the two comment types differ on that aspect.  
We thus split the source code artifacts from each comment, discard all the punctuation marks, and then determine the source token ratio over all tokens from the comment.  
Fig. \ref{fig:ce} shows box plots of source token ratio between both comment types. We see that median ratio of useful comments is relatively higher than that of non-useful comments.
More interestingly, we found the ratio difference statistically significant (\ie\ \emph{p-value=0.00$<$0.05, Cohen's D=0.17}) which is interesting. 
However, we further decompose our analysis, and our investigation with quartiles reports
that source token ratio is significantly higher for useful comments for all four quartiles with large effect size-- Q1 (\ie\ \emph{p-value=0.00$<$0.05, Cohen's D=0.61}), Q2 (\ie\ \emph{p-value=0.00$<$0.05, Cohen's D=0.70}), Q3 (\ie\ \emph{p-value=0.00$<$0.05, Cohen's D=0.92}) and Q4 (\ie\ \emph{p-value=0.00$<$0.05, Cohen's D=0.42}).
Finally, Kruskal Wallis test confirmed that usefulness of a review comment is significantly affected (\ie\ \emph{p-value=0.003}) by its source token ratio, which supports our initial conjecture.

\textbf{Conceptual Similarity:}
According to \citeauthor{useful}, usefulness of review comments is mostly determined based on their code change triggering capability in practice. 
Relevance of concepts (\eg\ concerns) discussed in the comment with the submitted code changes could be a proxy to such triggering capability.
Concept or feature location communities also hold a similar assumption in slightly different contexts \cite{refoqus}. 
We thus are interested to find out how relevant the review comments (\ie\ containing code elements and associated texts) are to the changed code submitted for review, and whether such relevance significantly affects their usefulness or not. In information retrieval, relevance or conceptual similarity between two text documents is generally determined based on their lexical similarity. 
Cosine similarity is a widely used metric to determine lexical similarity between two entities where each entity is considered as a vector of words \cite{antoniol}. We thus consider each in-line review comment and each changed source line as two individual entities, convert them into two vectors of terms by applying standard natural language preprocessing (\ie\ stop word removal, splitting), and then determine their cosine similarity.
Since the in-line review comment targets a particular source line, we treat each of the changed lines (\ie\ added, deleted, modified) as an individual entity during similarity calculation.
For each review comment, we then analyze all the changed source lines of the same file from earlier commits of the same pull request, and collect the maximum cosine measure. The measure is based on the shared vocabulary between the review comment and the changed source line, and thus, we consider the measure as a proxy to the relevance of the comment to the changed code. 


\begin{figure}[!t]
\centering
\includegraphics[width=2.5in ]{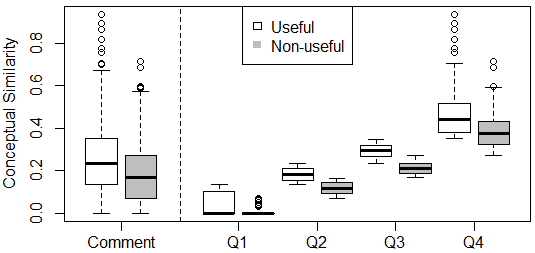}
\vspace{-.3cm}
\caption{Conceptual similarity between review comments and the changed code}
\label{fig:consim}
\vspace{-.7cm}
\end{figure}

From Fig. \ref{fig:consim}, we see that useful review comments are more relevant, \ie\ lexically or conceptually similar, to the changed code submitted for review. That is, they refer to the concepts embedded in the changed code more extensively than the non-useful comments, which is intuitive and also supports our initial conjecture. According to our analysis, useful comments have a median relevance measure of 0.24 which is 0.17 for non-useful comments.
Statistical tests report that the similarity measures of useful comments are significantly higher (\ie\ \emph{p-value=0.00}) with medium effect size (\ie\ \emph{Cohen's D=0.42}) than that of the non-useful comments. 
Further investigation using quartile analyses (Fig. \ref{fig:consim}) reported that
 relevance measures are also significantly higher for useful comments than the counterpart for each of the four quartiles--Q1 (\ie\ \emph{p-value=0.00, Cohen's D=0.91}), Q2 (\ie\ \emph{p-value=0.00, Cohen's D$>$1.00}), Q3 (\ie\ \emph{p-value=0.00, Cohen's D$>$1.00})  and Q4 (\ie\ \emph{p-value=0.00, Cohen's D=0.83}).


\begin{figure}[!t]
\centering
\includegraphics[width=2.4in ]{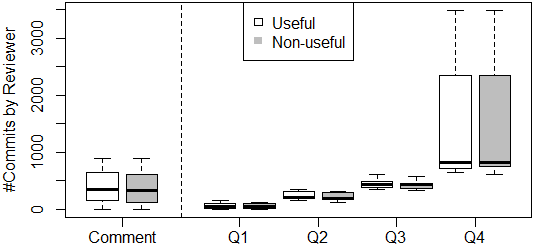}
\vspace{-.3cm}
\caption{Authorship of developers providing useful and non-useful review comments}
\label{fig:authorship}
\vspace{-.5cm}
\end{figure}

\textbf{Summary of the Findings--Answering RQ$\mathbf{_1}$:} Useful code review comments are significantly different than non-useful comments in terms of several textual properties of the comments such as \emph{code element ratio, stop word ratio} and \emph{conceptual similarity}, and not significantly different in terms of \emph{reading ease} and \emph{question ratio}.

\subsection{Developer Experience}\label{sec:devexp}
\textbf{Code Authorship:}\label{sec:authorship}
Code authorship is a traditional metric for estimating developer's experience which is reported to positively
influence the usefulness of code review comments \cite{useful,slowing}. 
According to \citeauthor{useful}, reviewers even with minimum authoring experience on a target source file (\ie\ changed once) can provide higher number of useful comments than the reviewers without any authoring experience.  
However, these findings were derived from the systems of \emph{Microsoft} of different application domains.  
We thus revisit their findings in our domain (\ie\ brand analytics), and investigate the correlation between code authorship and comment usefulness. 

We calculate the number of total commits submitted by the developers on a target file under review as well as on the whole subject system. Please note that only commits submitted before the date of the comment are considered during file level experience approximation.
We found that reviewers of 46.12\% of the useful comments changed the target file at least once whereas such statistic for non-useful comments is 49.00\%. The finding is a bit counter-intuitive, and it also does not support the earlier finding \cite{useful}.
However, our Kruskal-Wallis test on these statistics reported that comment usefulness is affected by the presence or absence of authorship experience of the reviewers (\ie\ \emph{p-value=0.00$<$0.05}). 
We then consider the file level commits of the reviewers of both comment types, and investigate their impact upon comment usefulness. 
We performed a Kruskal-Wallis test on file level commit counts for the reviewers, and found that their distributions are significantly different (\ie\ \emph{p-value=0.03$<$0.05}). This suggests that the usefulness of review comments is indeed influenced by the authorship of the reviewers, which supports the earlier finding \cite{useful}. We also repeat the investigation for commits on the whole system (\ie\ overall authoring experience), and found similar finding, \ie\ significant difference (\ie\ \emph{p-value=0.00$<$0.05}) in the distributions of commit count. 
Unfortunately, Mann-Whitney Wilcoxon test reported the opposite (\ie\ \emph{p-value=0.27$>$0.05, Cohen's D=0.12}).   
Thus, we further investigate, and divide the commit counts on whole system into four quartiles. 
Fig. \ref{fig:authorship} summarizes our findings from the investigation. Outliers are omitted from the plot to preserve visual clarity.
We performed Mann-Whitney Wilcoxon tests, and found that authorship (of the reviewers) difference  is significantly different between useful and non-useful comments for two quartiles--Q2 (\ie\ \emph{p-value=0.00$<$0.05, Cohen's D=0.42}) and Q3 (\ie\ \emph{p-value=0.74$>$0.05, Cohen's D=0.56}), and not significantly different for the rest two quartiles-- Q1 (\ie\ \emph{p-value=0.74$>$0.05, Cohen's D=0.08}) and Q4 (\ie\ \emph{p-value=0.80$>$0.05, Cohen's D=0.21}).
That is, too little or too much authoring experience on the subject system did not make a difference on the usefulness of review comments, \ie\ each of these experience levels produced similar amount of useful and non-useful comments.
On the contrary, medium level experience made a difference significantly. 
However, higher level authoring experience definitely helped the code reviewers to provide more useful review comments.


\begin{figure}[!t]
\centering
\includegraphics[width=2.4in ]{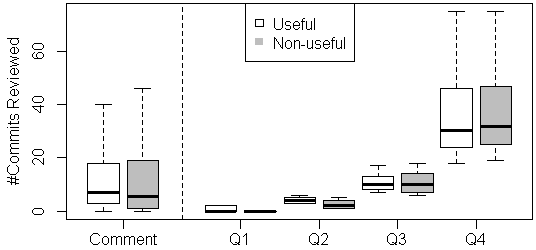}
\vspace{-.4cm}
\caption{Review experience of developers making useful and non-useful review comments}
\label{fig:reviewership}
\vspace{-.7cm}
\end{figure}

\textbf{Code Reviewership:}\label{sec:reviewership}
\citeauthor{useful} suggest a strong relationship between code reviewing experience of the developers and the usefulness of their provided review comments.
That is, experienced reviewers are likely to provide more useful comments than first time reviewers.
\citet{ownership} even suggest review activities to be considered as a metric for code ownership.
We revisit the earlier findings in the context of \emph{ABC Company}, and investigate how prior experience of the reviewers affect the usefulness of their comments. 

We determine review experience of the developers by calculating number of earlier commits on a target file currently under review, total commits, and pull requests reviewed by them from a subject system. 
We found that code reviewers reviewed the same file at least once for 87.22\% of the useful comments whereas such statistic for the non-useful comments is 84.54\%.
We also found that 86.25\% of the reviewers reviewed more than once for useful comments compared to 79.09\% for the non-useful comments. 
Although our findings do not strongly support the earlier findings, they align with them comfortably \cite{useful}.
When actual commits on the target file for each reviewer are considered, we also find that they are significantly different (\ie\ \emph{p-value=0.02$<$0.05}) for useful and non-useful review comments.
However, the effect size of their difference is small (\ie\ \emph{Cohen's D=0.05}). 
We thus perform Kruskal Wallis test on them, and found that comment usefulness is significantly affected (\ie\ \emph{p-value=0.00$<$0.05}) by the reviewed commit counts of the reviewers. 
We further investigate into this, and perform quartile analysis \cite{nontechnical}.
Fig. \ref{fig:reviewership} shows our comparative analysis between useful and non-useful comments. Outliers are omitted from the plot to preserve visual clarity.
We perform Mann-Whitney Wilcoxon tests on each of the quartiles, and found significant difference for Q1 (\ie\ \emph{p-value=0.00$<$0.05, Cohen's D=0.99}) and Q2 (\ie\ \emph{p-value=0.00$<$0.05, Cohen's D$>$1.00}), and not significant difference for Q3 (\ie\ \emph{p-value=0.72$>$0.05, Cohen's D=0.04}) and Q4 (\ie\ \emph{p-value=0.18$>$0.05, Cohen's D=0.27}). 
That is, reviewers with high level reviewing experience did not provide significantly higher number of useful comments than their non-useful comments.
On the contrary, reviewers with low to medium level experience did  exactly that. This might be partially explained from the earlier finding \cite{useful} that suggests that developers generally improve their reviewing experience during the \emph{``ramp up period"} (\ie\ first one-year of hire), and then the experience reaches into plateau.
However, according to our findings, increased reviewing experience on the target file definitely helps the reviewers provide more useful comments.
For example, reviewers with top level experience (\ie\ Q4) provided 160 useful review comments compared to 126 non-useful comments by them or 136 useful comments by the least experienced reviewers (\ie\ Q1). 


We also consider the total number of commits reviewed by the reviewers from a target system, and found that they are not significantly different (\ie\ \emph{p-value=0.63$>$0.05, Cohen's D=0.01}) for two comment types.
However, our Kruskal Wallis test reported that such review statistic has a significant influence (\ie\ \emph{p-value=0.00$<$0.05}) on the usefulness of review comments. 
We also observed similar findings when total number of pull requests reviewed by the reviewers is considered. 
We found that the statistic does not differ significantly (\ie\ \emph{p-value=0.33$>$0.05}) for useful and non-useful comments, but Kruskal Wallis test reported its significant influence (\ie\ \emph{p-value=0.00$<$0.05}) on comment usefulness. 
 

\begin{figure}[!t]
\centering
\includegraphics[width=2.5in ]{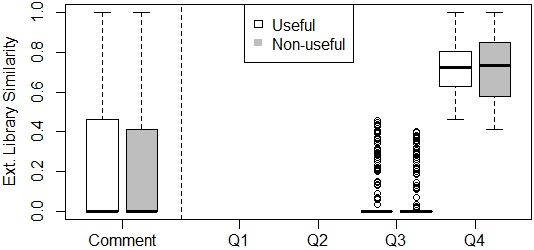}
\vspace{-.3cm}
\caption{Reviewer's experience on external libraries included in the file under review}
\label{fig:extlibsim}
\vspace{-.5cm}
\end{figure}

\textbf{External Library Experience:}\label{sec:libexp}
\citet{correct} suggest that experience with the external libraries included in a source file is an effective metric to determine the eligibility of a developer as the reviewer of that file.
That is, the developers who have working experience with the dependencies of a target file are likely to provide better reviews than the ones with no such experience \cite{correct}.  
We are interested to find out if such experience of the developers is connected to the usefulness of their review comments or not.
We thus determine similarity between the imported libraries in the target file (\ie\ attached to a review comment) and that from other files recently authored by the reviewer, and consider such similarity measure as a proxy to her experience with the external libraries. 
Then we contrast corresponding measures between useful and non-useful comments.

From Fig. \ref{fig:extlibsim}, we see that reviewers' minimum library experience is a bit higher for useful comments than that for non-useful review comments. 
However, our Mann-Whitney Wilcoxon test reported that their difference is not statistically significant (\ie\ \emph{p-value=0.29$>$0.05, Cohen's D=0.05}). 
Interestingly, a Kruskal Wallis test reported that experience with external libraries can significantly influence the usefulness of the review comments (\ie\ \emph{p-value=0.00$<$0.05}). 
We thus compare between two comment types using quartile analysis.
The median measure is 0 for both comments which leads the comparison to only Q3 and Q4.
We found that Q3 for useful comment is significantly higher (\ie\ \emph{p-value=0.03$<$0.05, Cohen's D=0.18}) than that for non-useful comments, which partially supports our initial conjecture about reviewers' experience on external libraries.  
For Q4, we did not find any significant difference (\ie\ \emph{p-value=0.62$>$0.05, Cohen's D=0.12}), and thus, does not support our conjecture.

\textbf{Summary of the Findings--Answering RQ$\mathbf{_2}$:} Prior experience with a file being reviewed in terms of authorship and reviewing activity of the developers can significantly influence the usefulness of their review comments. However, such experience of the developers and the usefulness of review comments
provided by them always do not have a linear relationship, which further supports the earlier findings \cite{useful}.      


\begin{figure}[!t]
\centering
\includegraphics[width=3.5in ]{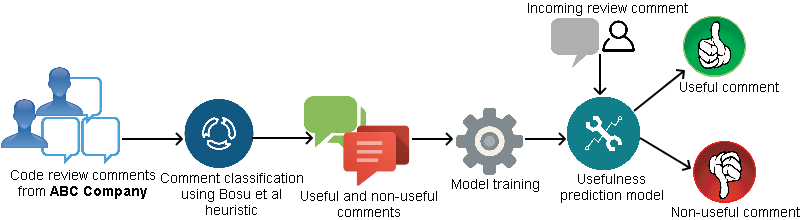}
\vspace{-.6cm}
\caption{Schematic diagram of the proposed prediction model--RevHelper}
\label{fig:revhelper}
\vspace{-.6cm}
\end{figure}

\section{RevHelper: Automatic Prediction Model for Review Comment Usefulness}
\label{sec:tagging}
Given a significant part (\ie\ 34.5\%) of the code review comments are non-useful \cite{useful}, 
an automatic technique is warranted for identifying and improving them before the review submission. 
Our comparative study reported that useful review comments significantly differ from non-useful review comments in terms of several textual properties,
and their reviewers have more experience.
We thus develop a model for automatic usefulness prediction (\ie\ \emph{useful} or \emph{non-useful}) of review comments      
by applying machine learning on these textual and experience based features.
In this section, we discuss different steps of our model development. Fig. \ref{fig:revhelper} shows the schematic diagram of our prediction model.

\subsection{Feature Calculation}\label{sec:featcalc}
We select a total of 15 independent variables (\ie\ several features yield multiple variables) from two dimensions--comment texts and developer experience-- and one response variable (\ie\ useful or non-useful) for our model.       
Although useful and non-useful comments do not vary significantly for some of the features according to the comparative study, we still wanted to investigate their predicting capabilities in a practical setting.
We thus apply all the features to our model.
While the textual features come from comment texts, the corresponding reviewers' experience is determined based on the analysis of version control history and code review history of the subject systems. 
Each of the comment instances is manually annotated as either \emph{useful} or \emph{non-useful} (details in Section \ref{sec:dataset}).
Less than 4\% of the samples in our dataset contain missing values for one or more features, and such values are either discarded or replaced with the mean of corresponding features.

\subsection{Development of the Usefulness Prediction Model}\label{sec:model}
Since the relationships between the usefulness of a review comment and corresponding features might be complex according to our comparative study, we choose three algorithms with different learning strategies--\emph{Random Forest} \cite{randomforest}, \emph{Logistic Regression} \cite{buse} and \emph{Naive Bayes} \cite{msrch2015masud}--
to train our classification model.
While Random Forest assumes the non-linear relationship between independent variables and the response variable, Logistic Regression does the opposite and returns binary outcomes.
On the other hand, Naive Bayes assumes the independence among the predictor variables, and determines the response based on the \emph{maximum log-likelihood} based training of the model parameters.
We apply 10-fold cross-validation for training and testing of the model when Naive Bayes and Logistic Regression are considered. That is, nine folds of our data are used for training and one fold is applied to testing at any moment, and this process iterates until all data are used for training and testing.
Then the results are averaged from the 10 models, and the final performance is reported.
On the other hand, Random Forest algorithm is robust enough to avoid over-fitting, and it does not warrant a cross-validation \cite{rfoverfit,randomforest}.
However, it randomly selects the predictor variables for tree generation, and thus, warrants multiple test runs.
We run our Random Forest based model 10 times, train with 65\% of the randomly sampled data (\ie\ different seed points) and a set of 2,000 decision trees, and then report the average test performance.  
We use machine learning workbench--WEKA \cite{weka}--and R--for training, testing and validation of our models.

 \begin{table*}[!t]
\centering
\caption{Performance of RevHelper}\label{table:result}
\vspace{-.2cm}
\resizebox{6.4in}{!}{%
\begin{threeparttable}
\begin{tabular}{l|l|l|c|c||c|c||c|c|c|c}
\hline
\multirow{2}{*}{\textbf{Learning Algorithm}} &  \multirow{2}{*}{\textbf{Dimension}} &  \multirow{2}{*}{\textbf{Feature Set}} & \multicolumn{2}{c||}{ \textbf{Useful Comments}} & \multicolumn{2}{c||}{\textbf{Non-useful Comments}} & \multicolumn{4}{c}{\textbf{All Comments}}\\
\hhline{~~~--------}
& & & \textbf{Precision} & \textbf{Recall} & \textbf{Precision} & \textbf{Recall} & \textbf{Precision} & \textbf{Recall} & \textbf{F$_1$-score}  & \textbf{Accuracy} \\
\hline
\hline
\multirow{2}{*}{Naive Bayes (NB)} & textual + & \{all features\} & 60.90\% & 57.10\% & 50.60\% & 54.40\% & 56.30\% & 55.90\% & 56.10\% &55.91\%\\
\hhline{~~---------}
& experience & \{all features\}$_{PCA}$ & 61.30\% & 66.00\% & 53.30\% & 48.20\% & 57.70\% & 58.10\% & 57.90\% & \textbf{58.06}\% \\
\hline
\hline
Logistic  & textual +  & \{all features\} & 60.70\% & 71.40\% & 54.60\% & 42.80\% & 58.00\% & 58.60\% & 58.30\% & \textbf{58.60}\%\\
\hhline{~~---------}
Regression (LR) & experience & \{all features\}$_{PCA}$ & 60.20\% & 72.80\% & 54.50\% & 40.40\% & 57.70\% & 58.30\% & 58.00\% & 58.33\%  \\
\hline
\hline
\multirow{4}{*}{Random Forest (RF)} & textual & \{RE, SWR, QR, CER, CS\}  & 60.52\% & 65.49\% & 50.49\% & 45.18\% & 56.04\% &  56.43\%  & 56.23\% & 56.58\%\\
\hhline{~----------}
 & experience & \{CA, CR, ELE\} & 66.73\% & \textbf{76.29}\% & 63.03\% & 51.29\% & 65.08\% & 65.13\% & 65.10\% & \textbf{65.37}\%\\
\hhline{~----------}
 & textual +  &  \{all features\} & 65.68 \% & 73.30\% & 59.73\% & 50.82\% & 63.02\% & 63.27\% & 63.14\% & \textbf{63.45}\% \\
\hhline{~~---------}
& experience & \{\textbf{CS, CA, CR, ELE, SWR}\} & \textbf{67.93}\% & 75.04\% & \textbf{63.06}\% & \textbf{54.54}\% & \textbf{65.76}\% & \textbf{65.89}\% & \textbf{65.82}\% & \textbf{66.06}\%\\
\hline

\end{tabular}
\centering
\end{threeparttable}
}
\vspace{-.5cm}
\end{table*}

\section{Experiment}\label{sec:experiment}
We conduct experiments using 1,482 code review comments from \emph{ABC Company} with 10-fold cross-validation tests and a separate case study.
Thus, we evaluate our model with real code review data from the commercial subject systems against a set of popular performance metrics.
In order to further validate our performance, we compare with three baseline comment classification models that were employed by \citet{useful}.
We thus answer three research questions as follows: 
\begin{itemize}
\item \textbf{RQ$\mathbf{_3}$}: How does the proposed model perform in predicting the usefulness of code review comments?
\item \textbf{RQ$\mathbf{_4}$}: How effective are the dimensions--comment texts and corresponding reviewers' experience--as a proxy to the usefulness of a review comment?
\item \textbf{RQ$\mathbf{_5}$}: Can RevHelper outperform the baseline models in predicting usefulness of the code review comments?
\end{itemize}


\subsection{Experimental Dataset}\label{sec:expdataset}
\textbf{Dataset Collection:} We conduct experiments using a total of 1,482 code review comments extracted from the review history of four commercial subject systems--CS, SM, MS and SR--of ABC Company.
While 1,116 of them were previously collected for comparative study (Section \ref{sec:dataset}), we extend the dataset for evaluation and validation following the same steps.

\textbf{Ground Truth Selection:} Each of our selected review comments is manually annotated based on the \emph{change triggering} heuristic of \citet{useful}.
The heuristic is objective and derived from the interviews with professional developers from the industry, and thus, our choice of the heuristic for ground truth selection is justified. 
That is, if the review comment triggers a code change to its proximity  (\ie\ 1--10 lines) in the subsequent commits, the comment is considered \emph{useful} and vice versa. Please check Section \ref{sec:dataset} for more detailed steps.

\textbf{Replication:} All experimental data and supporting materials
are hosted online \cite{revhelper} for replication or third-party reuse.

\begin{figure}[!t]
\centering
\includegraphics[width=2.8in ]{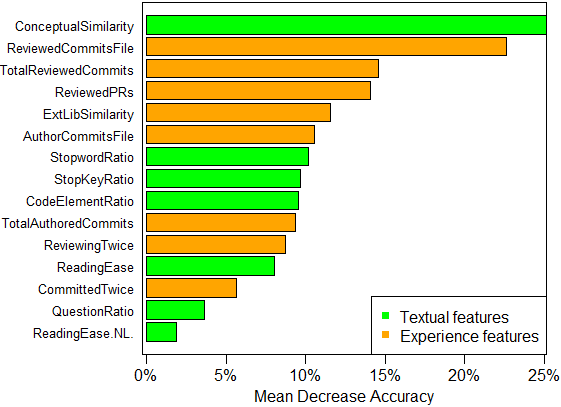}
\vspace{-.3cm}
\caption{Importance of 15 independent variables in RevHelper$_{RF}$}
\label{fig:importance}
\vspace{-.3cm}
\end{figure}

\begin{figure}[!t]
\centering
\includegraphics[width=3.5in ]{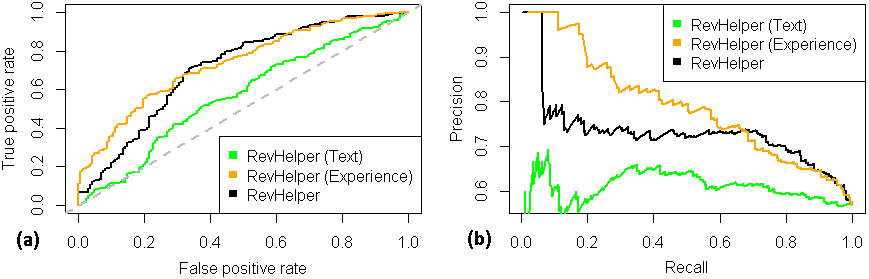}
\vspace{-.6cm}
\caption{(a) ROC and (b) Precision-Recall curves of RevHelper$_{RF}$}
\label{fig:roc}
\vspace{-.6cm}
\end{figure}

\subsection{Evaluation of RevHelper}\label{sec:evaluation}

\textbf{Performance of RevHelper--Answering RQ$\mathbf{_3}$:} 
We evaluate our prediction model with a dataset of 1,116 review comments using four appropriate performance metrics--\emph{precision}, \emph{recall}, \emph{F$_1$-score} and \emph{classification accuracy}.
We perform 10-fold cross-validation on Naive Bayes (NB) and Logistic regression (LR) based models and run Random Forest (RF) based model 10 times. Table \ref{table:result} summarizes our performance details.  

From Table \ref{table:result}, we see that our model predicts usefulness of a review comment with up to 66\% overall accuracy which is promising according to related literature on classification \cite{useful,buse,msrch2015masud}. Given that ours is the first automated support of its kind, such performance can make a difference during code review.
Our model exploits the non-linear relationships between comment usefulness and the independent variables using Random Forest learning algorithm, and provides better performance than the two other variants. For example, the model variants--RevHelper$_{NB}$ and RevHelper$_{LR}$--provide a maximum of 59\% overall accuracy even after applying certain feature optimization technique such as Principal Component Analysis (PCA). On the other hand, the final model--RevHelper$_{RF}$--provides 66\% overall accuracy with 66\% overall precision and 66\% overall recall. While decision tree based model might be less explanatory, feature importance graph (\eg\ Fig. \ref{fig:importance}) demonstrates the relative importance of the predictors of our model. 
Instant usefulness prediction during review submission along with insightful rationale behind the prediction (\eg\ details of predictor variables) is likely to help the reviewers improve their review comments. 
Especially, the non-useful comments lacking salient features (\eg\ relevant code elements) could be benefited from our instant supports.

Thus, to answer \textbf{RQ$\mathbf{\mathbf{_3}}$}, our model predicts usefulness of a review comment with 66\% overall accuracy, 66\% precision, recall and F$_1$-score which are promising. 


\textbf{Role of Dimensions--Answering RQ$\mathbf{_4}$:}
We consider two distinctive dimensions -- comment texts and reviewer's experience -- as a proxy to the usefulness of code review comments.
As investigation details in Table \ref{table:result} and performance curves in Figures \ref{fig:importance}, \ref{fig:roc}  suggest, experience of the reviewers is a better predictor of the usefulness of their comments than the textual characteristics of the comments. 
Unfortunately, prediction based on only reviewer's experience is of limited insights, less actionable and might not help much in improving the review comments in practice. Hence, a model that analyzes the actual content of a review comment and predicts instantly and reliably is warranted.
Thus, although the textual features have relatively less prediction capability and addition of this dimension to the model yields marginal improvement in performance,
our choice of introducing and combining both dimensions is possibly justified. Besides, from Fig. \ref{fig:importance}, we see that exclusion of one or more textual features (\eg\ \emph{conceptual similarity}) can significantly decrease the overall accuracy of the prediction model. 
This signals the importance of the textual dimension and makes it indispensable to the model.

Thus, to answer \textbf{RQ$\mathbf{_4}$}, reviewer's experience is more effective than textual content of a review comment in predicting usefulness of the comment. However, textual features could be more meaningful to the developers, and its addition to the model also marginally improves the performance. 


\subsection{Comparison with Baseline Models using Case Study}\label{sec:compare}
Although the above evaluation with several performance metrics and industrial data extensively analyzed the performance of our prediction model, we also wanted a comparison with the state-of-the-art. 
However, the existing model of \citet{useful} cannot predict the usefulness of a new review comment due to the nature of its learned features.
Many of their features are not available during the submission of a code review. Hence, comparing with \citet{useful} as is might be neither feasible nor fair.
Thus, we develop three variants of \citeauthor{useful} using a subset of their features--\emph{keyword count} and \emph{comment's sentiment}--only which are available during review comment submission like ours.
Please note that the omitted features from \citeauthor{useful} are not available during submission of a new review comment and thus, could not be used for its usefulness prediction.
Their methodologies, tools and learning algorithms are also carefully adopted. 
In particular, we used their keyword list \cite{useful}, sentiment analysis tool (\ie\ MSR-Splat \cite{splat}) and machine learning algorithm (\ie\ Classification and Regression Tree--CART \cite{cart}) to implement those variants. 
Then we prepare a separate validation set containing 366 code review comments totally unseen by any model, and compare our model with those variants using a case study.    

\begin{table*}[!t]
\centering
\caption{Comparison with Baseline Models}\label{table:compare}
\vspace{-.2cm}
\resizebox{7in}{!}{%
\begin{threeparttable}
\begin{tabular}{l|c|c|c|c|c|c||c|c|c|c|c|c||c}
\hline
\multirow{3}{*}{\textbf{Classifier}} & \multicolumn{6}{c||}{\textbf{Test Dataset}} &  \multicolumn{6}{c||}{\textbf{Validation Dataset}} & \textbf{Significance}  \\
\hhline{~-------------}
 &  \multicolumn{2}{c|}{ \textbf{Useful Comments}} & \multicolumn{2}{c|}{\textbf{Non-useful Comments}} & \multicolumn{2}{c||}{\textbf{All Comments}} & \multicolumn{2}{c|}{ \textbf{Useful Comments}} & \multicolumn{2}{c|}{\textbf{Non-useful Comments}} & \multicolumn{2}{c||}{\textbf{All Comments}} & \textbf{p-value} \& \\
\hhline{~------------}
& \textbf{Precision} & \textbf{Recall} & \textbf{Precision} & \textbf{Recall}& \textbf{F$_1$-score} & \textbf{Accuracy}  & \textbf{Precision} & \textbf{Recall} & \textbf{Precision} & \textbf{Recall} & \textbf{F$_1$-score} & \textbf{Accuracy} & \textbf{Cohen's D} \\
\hline
\citeauthor{useful}-I & 55.30\% & 98.50\% & 35.70\% & 1.00\% & 40.09\% &  55.02\% & 71.60\% & \textbf{100.00}\% & 0.00\% & 0.00\% & 0.00\% & \textbf{71.58}\% & --\\
\hline
\citeauthor{useful}-II-a & 55.40\% & \textbf{100.00}\% & 0.00\% & 0.00\% & 0.00\% &  55.38\% & 71.60\% & 100.00\% & 0.00\% & 0.00\% & 0.00\% & 71.58\% & -- \\
\hline
\citeauthor{useful}-II-b & 55.40\% & 95.10\% & 44.40\% & 4.80\% & 42.64\% & 54.84\%& 71.90\% & 98.50\% & 42.90\% & \textbf{2.90}\% & 61.05\% & 71.31\%& -- \\
\hline
\citeauthor{useful}-III-a & 55.00\% & 91.30\% & 40.70\% & 7.40\% & 43.60\%  & 53.85\%  & 71.60\% & 100.00\% & 0.00\% & 0.00\% & 0.00\% & 71.58\% & -- \\
\hline
\citeauthor{useful}-III-b & 56.80\% & 79.00\% & 49.40\% & \textbf{25.50}\% & \textbf{51.61}\% & 55.11\%  & 70.20\% & 77.50\% & \textbf{23.40}\% & \textbf{17.30}\% & 58.39\% & \textbf{60.38}\% & -- \\
\hline
\hline
RevHelper$_{txt}$ &60.52\% & 65.49\% & 50.49\% & 45.18\% & 56.12\%  & 56.58\% &  73.74\% & 71.40\% & 32.29\% & 34.90\% & 61.47\% & 61.15\% & 0.11$>$0.05 \& 0.22  \\
\hline
RevHelper$_{exp}$ & \textbf{66.73}\% & \textbf{76.29}\% & \textbf{63.03}\% & \textbf{51.29}\% & \textbf{64.66}\% & \textbf{65.37}\% &  \textbf{76.02}\% & 69.10\% & \textbf{35.86}\% & \textbf{44.20}\% & 63.07\% & 62.11\% & \textbf{*0.01}$<$0.05 \& 0.56  \\
\hline
RevHelper &  65.68\% & 73.30\% & 59.73\% & 50.82\% & 62.87\% & \textbf{63.45}\% & 74.09\% & 78.98\% & 35.27\% & 29.30\% & 63.83\% & \textbf{65.03}\% & \textbf{*0.00}$<$0.05 \& 0.46 \\
\hline


\end{tabular}
\centering
\textbf{a} = model uses comment sentiment group, \textbf{b} = model uses comment sentiment score, and \textbf{*}=significantly higher performance than the best baseline model--\citeauthor{useful}-III-b
\end{threeparttable}
}
\vspace{-.5cm}
\end{table*}

\begin{figure}[!t]
\centering
\includegraphics[width=1.65in ]{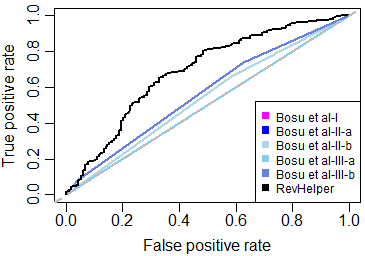}
\vspace{-.3cm}
\centering
\caption{Comparison of ROC with the baseline models}
\label{fig:compare-roc}
\vspace{-.4cm}
\end{figure}

\begin{figure}[!t]
\centering
\includegraphics[width=2in ]{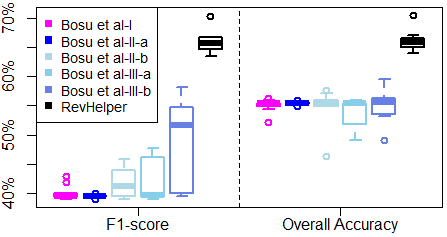}
\vspace{-.3cm}
\caption{Comparison of F$_1$-score and prediction accuracy with baseline models}
\label{fig:compare-box}
\vspace{-.6cm}
\end{figure}

Table \ref{table:compare} summarizes our comparative analyses using both test dataset (\ie\ cross-validation) and validation dataset.
We see that our model -- RevHelper -- performs significantly better than the baseline variants in terms of several performance metrics with both datasets.
The baseline models perform well for useful comments with about 70\% precision and up to 100\% recall. Unfortunately, they perform very poorly with non-useful comments. \citeauthor{useful}-I (\ie\ applies keyword counts) and \citeauthor{useful}-II-a (\ie\ applies comment sentiment group) 
returned 100\% false-positives for them.
Since our goal is to support developers in improving their review comments, especially the non-useful ones, performance with non-useful comments is of crucial importance.
The best performing baseline model-- \citeauthor{useful}-III-b (\ie\ applies both keyword counts and sentiment scores)--predicts the non-useful comments with 23\%--49\% precision and 17\%--26\% recall. On the other hand, our model provides 35\%--63\% precision and 29\%--51\% recall with an overall prediction accuracy of 65\%.
While our model has a significant error rate (35\%), it is still preferable to the existing alternatives given their poor performance (\ie\ 57\%--100\% false positives) with non-useful comments.
Besides, ours is the first model that predicts usefulness during a review submission.
The ROC curves in Fig. \ref{fig:compare-roc} 
also show that our model is more effective (\ie\ bigger AUC) than each of the baseline models. The box plots in Fig. \ref{fig:compare-box} also clearly demonstrate that our model performs significantly higher (\ie\ \emph{p-value=0.00, D$>$1.00}) than any of the baseline models.



Thus, to answer \textbf{RQ$\mathbf{_5}$}, our model --RevHelper-- outperforms the baseline models \cite{useful} significantly, and such findings clearly demonstrate a higher potential of our model for automatic developer support during code reviews.


\section{Threats to Validity}\label{sec:threat}
First, threats to internal validity relate to experimental errors and biases \cite{wordsim}. Our dataset selection involves manual analysis 
where the \emph{change triggering} heuristic of \citet{useful} was carefully applied.
While human errors cannot be avoided during analysis, we train each participant with pilot examples and conduct the analysis under constant supervision.
Besides, we perform random cross-check to avoid false-positives in the dataset. Finally, we collected a reasonable sample set of 1,482 review comments as an outcome of 30--35 man-hours work.


Second, the baseline model--\citeauthor{useful}-II-a-- returned 100\% false-positives for non-useful comments. Our investigation revealed that the suggested sentiment analyzer (\ie\ MSR-Splat \cite{splat}) found none of our comments  
\emph{negative} which is unlikely to be true.
To mitigate this threat, we repeated experiments using another parallel tool--Stanford Sentiment Analyzer \cite{sentiment}, but the overall validation findings did not change. 

Third, our study data are from one company and the systems are Python-based.
However, \citet{contempo} suggest that organizations and frameworks share common reviewing principles, and thus, our findings might also be generalizable.

\section{Related Work}\label{sec:related}

 Existing studies on code review explore various dynamics of the available code review practices \cite{whodoes,revgame,broadcast,convergent}, mine code review analytics \cite{devsee,olga-participation,nontechnical,slowing,shane-quality,fotuse-design,defective, ownership,expect} or review comment analytics \cite{panichella,useful,devsee,reduce}, and recommend appropriate reviewers for a code review \cite{reduce,pick,correct,kagdi,yu,rveffect,crase2016}.
While most of these studies perform qualitative or quantitative analysis to derive meaningful insights, 
there has been a marked lack in the application of such insights
to developing tools or techniques that can actually support the code review activities.
To the best of our knowledge, there exist not many tools or techniques that can automatically support a code review except the recently emerged code reviewer recommendation systems \cite{pick,correct,kagdi}.
In this work, we provide automated support in improving code review comments, and two existing studies--\citet{useful} and \citet{devsee}--are closely related to ours.
\citeauthor{useful} first study the perceptions of the professional developers from \emph{Microsoft} on the usefulness of code review comments, and identify several characteristics of useful and non-useful review comments from the qualitative study.  
 \citeauthor{devsee} conduct a similar study involving professional developers from open source domain--\emph{Mozilla}, and analyze the perceptions of the developers on the quality of code reviews.  
While both studies offer insights on the usefulness of review comments by employing qualitative methods, applicability of such insights (\ie\ using automated methods) is yet to be explored.
We harvest such learned insights on comment usefulness from them, and provide automated support to code reviews through a  
model for comment usefulness prediction. Such prediction could also be justified with meaningful rationale (\eg\ comment feature values).
Our automatic support is found promising and likely to help improve code review comments, especially the non-useful ones. Such exact support 
was not provided by any of the earlier studies.

\section{Conclusion and Future Work}\label{sec:conclusion}
To summarize, we propose a novel model for automatic and instant usefulness prediction of code review comments.
It exploits two dimensions--comment texts and developer experience--and uses machine learning for the prediction.
Our comparative study between useful and non-useful comments first motivates the model.
Experiments using 1,482 review comments from four commercial subject systems report 66\% overall prediction accuracy which is promising. 
Comparison with three variants of a baseline model using a case study and a validation set not only validates our empirical findings but also demonstrates the higher potential of our model. 
In future, we plan to extend our automatic supports for code reviews to personalized recommendation systems.

\textbf{Acknowledgement:} This research was supported in part by the Natural Sciences and Engineering Research Council of Canada (NSERC).

\balance

\bibliographystyle{plainnat}
\setlength{\bibsep}{0pt plus 0.3ex}
\bibliography{sigproc}  
%
%
\end{document}